\documentclass[11pt]{article}
\usepackage{latexsym}  
\usepackage{amssymb}
\usepackage{graphicx}
\usepackage{amsmath}

\begin{document}

\begin{center}

{\Large\bf U(3)-Flavor Nonet Scalar }\\
{\Large\bf as an Origin of the Flavor Mass Spectra}

\vspace{3mm}
{\bf Yoshio Koide}

{\it IHERP, Osaka University, 1-16 Machikaneyama, 
Toyonaka, Osaka 560-0043, Japan} \\
{\it E-mail address: koide@het.phys.sci.osaka-u.ac.jp}

\date{\today}
\end{center}

\vspace{3mm}
\begin{abstract}
According to an idea that the quark and lepton mass spectra 
originate in a VEV structure of a U(3)-flavor
nonet scalar $\Phi$, the mass spectra of the down-quarks and 
charged leptons are investigated.  The U(3) flavor symmetry is
spontaneously and completely broken by non-zero and non-degenerated
VEVs of $\Phi$, without passing any subgroup of U(3).
The ratios $(m_e+m_\mu+m_\tau)/(\sqrt{m_e}+\sqrt{m_\mu}+\sqrt{m_\tau})^2$
and $\sqrt{m_e m_\mu m_\tau}/(\sqrt{m_e}+\sqrt{m_\mu}+\sqrt{m_\tau})^3$
are investigated based on a toy model.
\end{abstract}

\vspace{3mm}
\noindent{\bf 1 \ Introduction}

The observed mass spectra of the quarks and leptons might provide 
a promising clue for a unified understanding of the quarks and 
leptons.   
In investigating an origin of the flavor mass spectra, 
we may expect that an approach based on symmetries will be promising.
However, when we want to introduce a flavor symmetry into our mass 
matrix model, we always encounter an obstacle, i.e.
a no-go theorem \cite{no-go} in flavor symmetries.
The theorem asserts that 
we cannot bring any flavor symmetry into a mass matrix model
as far as we consider a mass generation mechanism based on the
standard model.
The premises to derive the theorem are as follow:
(i) the SU(2)$_L$ symmetry is unbroken;
(ii) there is only one Higgs scalar in each sector 
(e.g. $H_u$ and $H_d$ for up- and down-quark sectors, respectively); 
(iii) 3 eigenvalues of $Y_f$ in each sector are non-zero and no-degenerate.
Therefore, we have three options \cite{no-go08} to evade this no-go theorem:
(A) a model with more than two Higgs scalars in each sector;
(B) a model with an explicit flavor symmetry breaking term; 
(C) a model with a new scalar whose vacuum expectation values (VEVs)
yield effective Yukawa coupling constants.
The approach (B) has adopted by many authors in phenomenological studies
of flavor symmetries.
However, we  want a model without such an explicit symmetry breaking  term.
The approach (A) induces a flavor-changing neutral current (FCNC) problem
\cite{FCNC}.  
In order to suppress the FCNC effects, we must make those Higgs scalars
heavy except for one of the linear combinations of those scalars.
However, it is not so easy to build such a reasonable suppression 
mechanism without an explicit symmetry breaking term.

Therefore, in the present paper,  we take a great interest in the option (C).
For example, we consider the following superpotential terms:
$$
W_{Y}= \sum_{i,j} \frac{y_u}{M}(Y_u)_{ij} {Q}_{i} H_u U_{j} 
+\sum_{i,j}\frac{y_d}{M} (Y_d)_{ij} {Q}_{i} H_d D_{j} 
$$
$$
+\sum_{i,j} \frac{y_\nu}{M}(Y_\nu)_{ij} {L}_{i} H_u N_{j} 
+\sum_{i,j}\frac{y_e}{M} (Y_e)_{ij} {L}_{i} H_d E_{j} 
+h.c. + y_R \sum_{i,j} N_i (M_R)_{ij} N_j ,
\eqno(1.1)
$$ 
where $Y_f$  ($f=u,d,\nu,e$) are not coupling constants, but U(3)-flavor
nonet fields \cite{Koide90,YK-Tanimoto96,YK-JHEP07}, 
and $Q$ and $L$ are quark and lepton 
SU(2)$_L$ doublet fields, respectively, and $U$, $D$, $N$, and $E$ are 
SU(2)$_L$ singlet matter fields.
The mass parameter $M$ denotes an energy scale of the effective theory.
For example, 
if we assume the following terms for an additional U(3)-nonet (gauge 
singlet) scalar 
$\Phi$ in the superpotential
$$
W_\Phi = \lambda_\phi {\rm Tr}[\Phi\Phi\Phi] +m_{\phi\phi}{\rm Tr}[\Phi\Phi]
+\mu_\phi^2 {\rm Tr}[\Phi]
+\lambda_{e} {\rm Tr}[\Phi\Phi Y_e] +m_{ee} {\rm Tr}[Y_e Y_e],
\eqno(1.2)
$$
we can obtain relations
$$
3 \lambda_\phi \Phi\Phi + 2 m_{\phi\phi}\Phi +\mu_\phi^2 {\bf 1}
+\lambda_e (\Phi Y_e+Y_e\Phi) = 0 ,
\eqno(1.3)
$$
$$
\lambda_e \Phi\Phi+ 2 m_{ee} Y_e = 0 ,
\eqno(1.4)
$$
from SUSY vacuum conditions $\partial W/\partial \Phi=0$
and $\partial W/\partial Y_e=0$, respectively.
(Here, for simplicity, we have drop the contribution
form $W_Y$.)
Therefore, we can obtain a bilinear mass relation for the
charged leptons
$\langle Y_e\rangle = -({\lambda_e}/{2 m_{ee}}) \langle\Phi\rangle 
\langle\Phi\rangle$, from Eq.(1.4).
This is entirely favorable for charged lepton mass relation 
as we state later.
By eliminating $Y_e$ from Eq.(1.3) with Eq.(1.4), we obtain
$$
c_3 \Phi\Phi\Phi + c_2 \Phi\Phi + c_1 \Phi + c_0 {\bf 1} =0 ,
\eqno(1.5)
$$
where 
$c_3={\lambda_e^2}/{m_{ee}}$,  $c_2 = -3\lambda_\phi$, 
$c_1 = - 2m_{\phi\phi}$ and $c_0 = -\mu_\phi^2$.
Thus, if we give values of the coefficients $c_n$  ($n=3,2,1,0$),
we can completely determine three eigenvalues of $\langle\Phi\rangle$,
so that we can give a charged lepton mass spectrum from Eq.(1.5).
Especially, it is worthwhile noticing that
a relation between 
${\rm Tr}[\Phi\Phi]$ and ${\rm Tr}^2[\Phi]$ is described by
${\rm Tr}[\Phi\Phi]/{\rm Tr}^2[\Phi] = 1-2 {c_1 c_3}/{c_2^2}$,
and  a ratio ${\rm det}\Phi/{\rm Tr}^3[\Phi]$ is given by
${\rm det}\Phi/{\rm Tr}^3[\Phi] = {c_0}{c_3}^2/c_2^3$.
We should note that the superpotential (1.2) does not include 
any explicit flavor symmetry breaking parameter.
The most distinctive feature of the present model is 
that the U(3) flavor symmetry is spontaneously and completely 
broken by the non-zero and non-degenerate VEVs of 
$\langle\Phi\rangle$, 
without passing any subgroup of U(3). 
(For example, differently from the present model,
a U(3)-nonet scalar $\Phi$ in Ref.\cite{YK-JHEP07} is
broken, not directly, but via a discrete symmetry S$_4$.)

The idea mentioned above is very attractive, because 
the model does not include conventional Yukawa coupling constants
which explicitly break the flavor symmetry.
However, a straightforward application of the model (1.2)
needs, at least, four different $\Phi$ fields, i.e.
$\Phi_f$ ($f=u,d,\nu,e$), because we know that mass spectra
in the four sectors  are  completely different from each 
other.
From the economical point of view in a unification model 
of quarks and leptons,  
we will consider that the $\Phi$ fields are, at most, two, i.e.
$\Phi_u$ and $\Phi_d$, which couple to the up-quark and
neutrino sectors and to the down-quark and charged lepton
sectors, respectively.

In the present paper, at the outset, we will begin
to investigate the down-quark and charged lepton sectors.
In the next section, Sec.2, we give a framework of the model,
and we will investigate the relations for 
${\rm Tr}[\Phi\Phi]/{\rm Tr}^2[\Phi]$ and 
${\rm det}\Phi/{\rm Tr}^3[\Phi]$, and also discuss a relation
between the down-quark masses $m_{di}$ and the charged lepton
masses $m_{ei}$.  
Those relations are essentially described by four parameters.
In Sec.3, we will give a speculation in order to obtain 
explicit values of those parameters,
although it is only a toy mode and it should not be seriously taken.
Finally, Sec.4 will be devoted to concluding remarks. 
 

\vspace{3mm}
\noindent{\bf 2 \ Model}

We assume
the following superpotential:
$$
W_\Phi = \lambda_\phi {\rm Tr}[\Phi\Phi\Phi] +m_{\phi\phi}{\rm Tr}[\Phi\Phi]
+\mu_\phi^2 {\rm Tr}[\Phi]
+\lambda_e {\rm Tr}[\Phi\Phi Y_e] +m_{ee} {\rm Tr}[Y_e Y_e]
$$
$$
+\lambda_d {\rm Tr}[\Phi\Phi Y_d] +m_{dd} {\rm Tr}[Y_d Y_d]
+m_{d\phi} {\rm Tr}[Y_d \Phi] ,
\eqno(2.1)
$$
where $\Phi$, $Y_e$ and $Y_d$ are U(3)-flavor nonet superfields
(for convenience, we denote $\Phi_d$ as $\Phi$ simply),
and the $m_{d\phi}$-term has been added in order to give a 
down-quark mass formula as we state later.
In order to couple $Y_e$ and $Y_d$ with the charged lepton
sector $LEH_d$ and down-quark sector $QDH_d$, respectively,
for example, we may assign additional U(1) charges $(q_e, -q_e)$
and $(q_d, -q_d)$ to the fields $(Y_e, E)$ and $(Y_d, D)$,
respectively.
However, such U(1) charges cannot be conserved in $W_\Phi$
unless the U(1) charges are also suitably assigned to the
coefficients in $W_\Phi$.
For the moment, we assume such phenomenological assignments of
the U(1) charges to the coefficients in $W_\Phi$.
 
From the SUSY vacuum conditions, 
we obtain
$$
\frac{\partial W}{\partial \Phi}=0 =
3\lambda_\phi \Phi\Phi +2 m_{\phi\phi} \Phi +\mu_\phi^2 {\bf 1} 
+\lambda_e (\Phi Y_e + Y_e \Phi)
+\lambda_{d} (\Phi Y_d + Y_d \Phi) + m_{d\phi} Y_d ,
\eqno(2.2)
$$
$$
\frac{\partial W}{\partial Y_e}=0 =
\lambda _e \Phi\Phi + 2 m_{ee} Y_e ,
\eqno(2.3)
$$
$$
\frac{\partial W}{\partial Y_d}=0 =
\lambda _d \Phi\Phi + 2 m_{dd} Y_d + m_{d\phi} \Phi ,
\eqno(2.4)
$$
so that we obtain the following relations
$$
\langle Y_e\rangle = -\frac{\lambda_e}{2 m_{ee}} \langle\Phi\rangle 
\langle\Phi\rangle ,
\eqno(2.5)
$$
$$
\langle Y_d\rangle = -\frac{\lambda_d}{2 m_{dd}}
\left( \langle\Phi\rangle \langle\Phi\rangle 
+\frac{m_{d\phi}}{\lambda_d} \langle\Phi\rangle \right) .
\eqno(2.6)
$$
(Hereafter, for simplicity, we will denote $\langle\Phi\rangle $,
$\langle Y_e\rangle$ and $\langle Y_d\rangle$ as
$\Phi$, $Y_e$ and $Y_d$.)
Eqs.(2.5) and (2.6) mean that the charged lepton masses $m_{ei}$ 
and down-quark masses $m_{di}$ are given by  
$$
m_{ei} = m_0^e z_i^2 ,
\eqno(2.7)
$$
$$
m_{di} = m_0^d (z_i^2 + \eta z_i) ,
\eqno(2.8)
$$
respectively, where $z_i = v_i/\sqrt{v_i^2+v_2^2+v_3^2}$ and
$v_i \delta_{ij} =\langle(\Phi_d)_{ij}\rangle$ in the diagonal
basis of $\langle \Phi_d\rangle$, so that the paremeter $\eta$
is given by
$$
\eta = \frac{m_{d\phi}/\lambda_d}{\sqrt{{\rm Tr}[\Phi\Phi]}} .
\eqno(2.9)
$$
The values of the quark mass ratios 
 $m_d/m_s=0.050$ and $m_s/m_b=0.031$ at $\mu=M_Z$  \cite{FK-qmass}
lead to $\eta\simeq -0.11$ and $\eta\simeq -0.13$, respectively,
so that we can understand the observed ratios by taking 
$\eta\simeq -0.12$ within one sigma deviation.\footnote{
However, this possibility is still controversial. 
Recent updated quark mass estimates  \cite{Ross-qmass07}
and \cite{Xing-qmass07} have
reported $(m_d/m_s=0.051; m_s/m_b=0.019)$  and
$(m_d/m_s=0.052; m_s/m_b=0.017)$, respectively, as the values  at
$\mu=M_{GUT} = 2\times 10^{16}$ GeV with $\tan\beta=10$. 
Although the both values of $m_d/m_s$ lead to $\eta \simeq -0.11$,
the values $m_s/m_b = 0.019$ and $m_s/m_b = 0.017$
lead to $\eta \simeq -0.17$ and $\eta \simeq -0.18$, respectively,
so that the unified description based on Eq.(2.8) fails.
However, note that the quark mass values are highly dependent
on the value of $\tan\beta$.
Besides, we do not always consider that the relations (2.7) and (2.8) 
are given at $\mu=M_{GUT}$.
The mass ratio $m_s/m_b$ is highly dependent on the
energy scale $\mu$.
Therefore, in the present paper, by considering that the scenario (2.8)
is applicable to the observed quark masses, we will go on investigating.
} 
(Here, in estimating the value $\eta$, we have used the values $z_i$ 
which are obtained from the pole mass values of the charged leptons,
because the ratios are insensitive to the energy scale $\mu$.)
This seems to offer a new view of the unified understanding of the
quark and lepton masses and mixings.

By substituting Eqs.(2.5) and (2.6) into (2.2), we again obtain the
same equation with (1.5), 
$$
c_3 \Phi\Phi\Phi + c_2 \Phi\Phi + c_1 \Phi + c_0 {\bf 1} =0 ,
\eqno(2.10)
$$
where
$$
c_3=\frac{\lambda_e^2}{m_{ee}} + \frac{\lambda_d^2}{m_{dd}} ,
\eqno(2.11)
$$
$$ 
c_2 = -3 \lambda_\phi \left( 1-\frac{1}{2}\frac{m_{d\phi}}{m_{dd}}
\frac{\lambda_d}{\lambda_\phi} \right) ,
\eqno(2.12)
$$
$$
c_1 = -2 m_{\phi\phi} \left( 1-\frac{1}{4}
\frac{(m_{d\phi})^2}{m_{\phi\phi} m_{dd}}  \right) ,
\eqno(2.13)
$$
$$
c_0 = -\mu_\phi^2 ,
\eqno(2.14)
$$
and the coefficients $c_n$ have the following relations with ${\rm Tr}[\Phi]$, 
${\rm Tr}[\Phi\Phi]$ and ${\rm det}\Phi$
$$
\frac{c_2}{c_3}= -{\rm Tr}[\Phi]  ,
\eqno(2.15)
$$
$$
\frac{c_1}{c_3}= \frac{1}{2} \left( 
{\rm Tr}^2[\Phi] -{\rm Tr}[\Phi\Phi] \right) ,
\eqno(2.16)
$$
$$
\frac{c_0}{c_3}= -{\rm det}\Phi .
\eqno(2.17)
$$ 
Here, it is convenient to define
the following parameters:
$$
\tilde{m}_{\phi\phi} = \frac{m_{\phi\phi}}{\lambda_\phi^2} , \ \ 
\tilde{m}_{ee} = \frac{m_{ee}}{\lambda_e^2} , \ \ 
\tilde{m}_{dd} = \frac{m_{dd}}{\lambda_d^2} , \ \ 
\tilde{m}_{d\phi} = \frac{m_{d\phi}}{\lambda_\phi \lambda_d} , \ \ 
\tilde{\mu}_\phi^2 = \frac{\mu_\phi^2}{\lambda_\phi^3} ,
\eqno(2.18)
$$
$$
\alpha =\frac{\tilde{m}_{dd}}{\tilde{m}_{ee}} , \ \ 
\beta =\frac{\tilde{m}_{\phi\phi}}{\tilde{m}_{dd}} , \ \ 
\gamma =\frac{\tilde{m}_{d\phi}}{\tilde{m}_{dd}} , \ \ 
\delta =\frac{\tilde{\mu}_\phi^2}{\tilde{m}_{dd}^2} . 
\eqno(2.19)
$$
Note that those parameters $\alpha$, $\beta$,  $\gamma$ 
and $\delta$ are invariant
under the scale transformation of the fields 
$\Phi\rightarrow \xi_\phi \Phi$,  $Y_e\rightarrow \xi_e Y_e$
and $Y_d\rightarrow \xi_d Y_d$, because 
$\tilde{m}_{\phi\phi}\rightarrow \tilde{m}_{\phi\phi}/\xi_\phi^4$, 
$\tilde{m}_{ee}\rightarrow \tilde{m}_{ee}/\xi_\phi^4$, 
$\tilde{m}_{dd}\rightarrow \tilde{m}_{dd}/\xi_\phi^4$,
$\tilde{m}_{d\phi}\rightarrow \tilde{m}_{d\phi}/\xi_\phi^4$ and
$\tilde{\mu}^2\rightarrow \tilde{\mu}^2/\xi_\phi^8$.
From Eqs.(2.15) and (2.16), we obtain
$$
v_1+v_2+v_3={\rm Tr}[\Phi]= \frac{3}{2} 
\lambda \tilde{m}_{dd} \frac{2-\gamma}{1+\alpha} ,
\eqno(2.20)
$$ 
$$
R\equiv \frac{v_1^2+v_2^2+v_3^2}{(v_1+v_2+v_3)^2}=
\frac{{\rm Tr}[\Phi\Phi]}{{\rm Tr}^2[\Phi]} =
1-2 \frac{c_1 c_3}{c_2^2} =
1 -\frac{4}{9}(1+\alpha) \frac{\gamma^2-4\beta}{
(\gamma-2)^2} .
\eqno(2.21)
$$
The deviation from the bilinear form in the down-quark mass formula 
[$\eta$ in the expression (2.8)] are described by 
$$
\eta = - \frac{2}{3} (1+\alpha) \frac{\gamma}{\gamma -2}
\left[ 1 -\frac{4}{9}(1+\alpha) \frac{\gamma^2-4\beta}{
(\gamma-2)^2} \right]^{-1/2} ,
\eqno(2.22)
$$
from Eq.(2.9).
On the other hand, the ratio ${\rm det}\Phi/{\rm Tr}^3[\Phi]$
is given by
$$
r_{123} \equiv  \frac{v_1 v_2 v_3}{(v_1+v_2+v_3)^3}=
\frac{{\rm det}\Phi}{{\rm Tr}^3[\Phi]} =
\frac{c_0 c_3^2}{c_2^3} = \frac{8}{27} \frac{(1+\alpha)^2}{(2-\gamma)^3}
\delta ,
\eqno(2.23)
$$
from (2.17).

From Eqs.(1.1), (2.5), (2.6) and (2.20), we can obtain $m_0^f$ ($f=e,d$) 
defined in Eqs.(2.7) and (2.8) as follows:
$$
m_0^f= -\frac{y_f}{M} \frac{\lambda_f}{2 m_{ff}} v^2 v_H
=-\frac{3}{8} \frac{y_f}{M} \frac{\lambda_f}{m_{ff}}
\lambda_\phi^2 \left( \frac{m_{dd}}{\lambda_d^2}\right)^2
\left(\frac{2-\gamma}{1+\alpha} \right)^2 
\frac{1}{(z_3+z_2+z_1)^2} v_H ,
\eqno(2.24)
$$
where $\langle \Phi_{ii}\rangle = v_i = v z_i$
($v^2=v_1^2+v_2^2+v_3^2$) and $v_H=\langle H_d^0\rangle$.
Since the order of $m_0^f$ is $m_0^f \sim m_{dd}^2/M m_{ff}$,
we can consider $m_{ff}/M \sim 10^{-2}$.  

In order to give explicit values of the mass spectra, we need
further assumptions. 
In the present paper, we are interested in the ratio $R$
which is given by (2.21),
because if we can give $R=2/3$,
which means VEV relation
$$
v_1^2+v_2^2+v_3^2 =\frac{2}{3} (v_1+v_2+v_3)^2 ,
\eqno(2.25)
$$
we can obtain the following mass relation \cite{Koidemass} 
for the charged leptons 
$$
m_e+m_\mu +m_\tau =\frac{2}{3}\left( \sqrt{m_e} +\sqrt{m_\mu}
+\sqrt{m_\tau} \right)^2 ,
\eqno(2.26)
$$
from the bilinear mass formula (2.5) on the diagonal basis of 
$\langle \Phi\rangle$.

We consider that the ratio $R$ is a fundamental quantity in the present model, 
so that we expect  
that the ratio $R$ will be expressed by a simple form.
Since the $m_{d\phi}$-term is an extra term from the point of view of 
an $e \leftrightarrow d$ symmetry in the superpotential (2.1), 
we consider on trial that the ratio $R$ will be independent 
of such a parameter $\gamma=\tilde{m}_{d\phi}/\tilde{m}_{dd}$.
This demands
$$
R = \lim_{\gamma\rightarrow \infty} R = \lim_{\gamma\rightarrow 0} R .
\eqno(2.27)
$$
From the requirement (2.27) for the case 
$\gamma \rightarrow \infty$, 
we obtain the relation
$$
\gamma= \beta +1 ,
\eqno(2.28)
$$
i.e.
$$
\tilde{m}_{d\phi} = \tilde{m}_{dd} +\tilde{m}_{\phi\phi} .
\eqno(2.29)
$$
Then, we get a simple expression
$$
R = 1-\frac{4}{9} (1+\alpha) .
\eqno(2.30)
$$
(For $\gamma\rightarrow 0$, we take $\beta \rightarrow -1$
from the condition (2.28), so that we again obtain the result (2.30).)
Note that the ratio $R$ is described only by one parameter 
$\alpha=\tilde{m}_{dd}/\tilde{m}_{ee}$.


\vspace{3mm}
\noindent{\bf 3 \ Speculations}

In the present section, in order to speculate the values of 
the parameters $\alpha$ and $\beta$, let us put the following 
assumptions on trial:
$$
\lambda_e +\lambda_\phi + \lambda_d =0 ,
\eqno(3.1)
$$
$$
m_{ee} + m_{\phi\phi} +m_{dd} + m_{d\phi} =0 ,
\eqno(3.2)
$$
although there is no theoretical ground for such requirements.
We consider that the ratio $R$ is a fundamental quantity in
the model, so that it is likely that the ratio is rational.
Therefore, we consider that the relations (3.1) and (3.2) are 
also rational.
Since the $m_{d\phi}$-term is concerned with $\Phi$ and $Y_d$,  
we consider that more fundamental parameter will be $\lambda_e$
rather than $\lambda_\phi$.
Therefore, we assume that the relation (3.1) will be expressed 
rationally in the unit of $\lambda_e$, e.g.
$$
\lambda_\phi = n \lambda_e , \ \ \  \lambda_d =-(n+1)\lambda_e ,
\ \ \ (n=1,2,\cdots).
\eqno(3.3)
$$
For the relation (3.2) with an additional assumption
$m_{d\phi} = m_{dd} + m_{\phi\phi}$ (cf. Eq.(2.29)), i.e. for 
$m_{ee} + 2m_{\phi\phi} + 2 m_{dd} =0$, 
we assume requirements similar to (3.3):
$$
2m_{\phi\phi} = n m_{ee} , \ \ \  2m_{dd} =-(n+1) m_{ee} , 
\ \ \ (n=1,2,\cdots).
\eqno(3.4)
$$
Since we define ${\rm Tr}[\Phi] >0$ and we search the solutions
with $\eta <0$, the signs of $m_{dd}$ and $m_{\phi\phi}$ must be opposite each other.
By considering the relation $m_{dd}=-(n+1) m_{\phi\phi}/n$ from (3.4), 
we must take $n$  as $n >0$. 
Then, the parameters $\alpha$ and $\beta$ are given by
$$
\alpha=\frac{m_{dd} \lambda_e^2}{m_{ee} \lambda_d^2}=-\frac{1}{2(n+1)} ,
\ \ \ 
\beta=\frac{m_{\phi\phi} \lambda_{d}^2}{m_{dd} \lambda_\phi^2}=-\frac{n+1}{n} ,
\eqno(3.5)
$$
and the ratio $R$ and parameter $\eta$ are given by
$$
R= \frac{5n+7}{9(n+1)} ,
\eqno(3.6)
$$
$$
\eta = - \frac{1}{\sqrt{(n+1)(5n+7)}} .
\eqno(3.7)
$$
We assume that the ratio $R$ should be as large as possible.
This demands $n=1$ in Eq.(3.6) with $n=1,2,\cdots$.
(On the other hand, the case $n=1$ gives a minimum of $|\eta|$.) 
Then, we can obtain $\alpha=-1/4$ which gives
the desirable relation 
$$
R=\frac{m_e+m_\mu+m_\tau}{(\sqrt{m}_e+\sqrt{m}_\mu+\sqrt{m}_\tau)^2}
= \frac{2}{3} .
\eqno(3.8)
$$
(However, this does not mean that we have derived the formula (3.8),
because the present scenario described in (3.1) - (3.4) have no
theoretical basis.
The choice $\alpha=-1/4$ is merely one of possible choices. ) 
If we accept the present speculation (3.7) with $n=1$,
we obtain a value of $\eta$
$$
\eta =-\frac{1}{\sqrt{24}} =-0.204 .
\eqno(3.9)
$$
Regrettably, 
the magnitude of the predicted value (3.9) is somewhat larger than 
the desirable value $\eta \simeq -0.12$ which is estimated from 
the formula (2.8) with the observed quark mass ratios \cite{FK-qmass}.
However, we do not consider that this discrepancy (3.9) in the 
present toy model denies the basic idea suggested in
Sec.2.\footnote{
In order to adjust the predicted values of $m_{di}$, 
for example, we may add a tadpole term $\mu_d^2 {\rm Tr}[Y_d]$
to the superpotential (2.1).
Then, the down-quark mass spectrum will be given by
$m_{di} = m_0^d (z_i^2 +\eta_2 z_i +\eta_0)$ instead of (2.8).
However, such an additional term will affect the coefficient $c_1$
defined in Eq.(2.10).
From the point of view of simplicity, in the present paper,
we do not consider such a modification.
} 
Rather, we consider that the order of the value (3.9) is reasonable,
so that our direction is basically right.

Finally, let us speculate the value of the ratio $r_{123}$.
From Eq.(2.23), we obtain
$$
r_{123} = 
\frac{\sqrt{m_e m_\mu m_\tau}}{(\sqrt{m}_e+\sqrt{m}_\mu+\sqrt{m}_\tau)^3}
= \frac{2}{27} \frac{n^3}{(n+1)^2(2 n+1)} \delta .
\eqno(3.10)
$$
Since the value $r_{123}$ should be zero in the limit $m_e\rightarrow 0$, 
we expect that the value is realized as small as possible.
This again demands $n=1$ in (3.10), and we obtain
$$
r_{123}=\frac{1}{2\cdot 3^4}\delta .
\eqno(3.11)
$$
Previously, we have assumed the constraints (3.1) and (3.2) for
the quadratic and cubic terms in the superpotential
(2.1), while, for the tadpole term, since the tadpole term is 
the $\mu_\phi^2$-term alone, we cannot put such a speculative
relation, so that we cannot speculate a value of $\delta$.
The value $\delta$ is completely free, although we consider
that the value is also rational.
Therefore, we give up the prediction of the value $r_{123}$, and
instead, we estimate of a value of $\delta$ from the
observed charged lepton mass rations. 
The observed charged lepton masses \cite{PDG06} give 
$$
z_1 = 0.01647, \ \ z_2= 0.23687, \ \ z_3 =0.97140 ,
\eqno(3.12)
$$
as the values of $z_i =\sqrt{m_{ei}/(m_e + m_\mu + m_\tau)}$.
Although we know that above values are excellently satisfy 
the relation (2.26) [i.e. (3.8)], the ``masses" in the present
model mean not ``pole" masses, but the ``running" masses.
For example, if we adopt the mass values \cite{Xing-qmass07}
at $\mu=2\times 10^{16}$ GeV which are estimated from a SUSY 
scenario with $\tan\beta=10$, we obtain
$$
z_1 = 0.01619, \ \ z_2= 0.23517, \ \ z_3 =0.97182 .
\eqno(3.13)
$$
The values (3.12) and (3.13) well satisfy the relation (3.8),
i.e.  within the deviation $2\times 10^{-6}$
and $3\times 10^{-3}$, respectively. 
However, for the ratio $r_{123}=z_1 z_2 z_3/(z_1+z_2+z_3)^3$, 
both values give slightly different values of $r_{123}$:
the values (3.12) give $r_{123}= 0.002063$, while the values (3.13)
give $ r_{123}=0.002013$.
(Besides, the value $r_{123}$ is considerably dependent on the 
value of $\tan\beta$.)
In the present paper, we ignore such a small difference.
Since we consider the parameter $\delta$ will be also expressed
with a concise rational value, we take it on trial as 
$$
\delta =\frac{1}{3} .
\eqno(3.14)
$$ 
Then, we obtain
$$
r_{123}=\frac{\sqrt{m_e m_\mu m_\tau}}{(\sqrt{m}_e+\sqrt{m}_\mu
+\sqrt{m}_\tau)^3} = 
\frac{1}{2\cdot 3^5} = 0.002058 .
\eqno(3.15)
$$
so that we obtain the predicted values of $z_i$, $z_1=0.01642$, 
$z_2=0.2369$ and $z_3 =0.97139$, 
which are in good agreement with (3.12) [and also 
(3.13)].
The rational value $\delta=1/3$ is plausible,
although we have no theoretical ground for $\delta=1/3$.


\vspace{3mm}
\noindent{\bf 4 \ Concluding remarks}

In conclusion, we have investigated the charged lepton and down-quark
mass spectra on the basis of a model in which the quark and lepton
mass spectra originate not in structures of Yukawa coupling 
constants, but in structures of VEVs of U(3)-flavor nonet 
(gauge singlet) fields $\Phi_u$ and $\Phi_d$.
We have proposed a mechanism which gives a bilinear form 
$M_e \propto \langle\Phi_d\rangle\langle\Phi_d\rangle$  
for the charged lepton mass matrix $M_e$,
and which gives a form $M_d \propto \langle\Phi_d\rangle
\langle\Phi_d\rangle + c \langle\Phi_d\rangle$ for
the down-quark mass matrix $M_d$.
The U(3)-flavor symmetry is spontaneously and completely broken
without passing any subgroup of U(3), i.e. directly. 
The VEV spectrum $\langle\Phi_d\rangle= v\, {\rm diag}(
z_1, z_2, z_3)$ is completely determined by the coefficients 
in the superpotential $W_\Phi$, (2.1).
The superpotential (2.1) does not include any symmetry breaking
term.
As shown in Sec.2, the VEV spectrum of $\langle\Phi_d\rangle$
(we have denoted $\Phi_d$ as $\Phi$ simply) 
is essentially described by four parameters which have been
defined in (2.19).
Thus, the VEV spectrum is closely related to the both parameters
in the charged lepton and down-quark sectors.
The rations $R=(m_e+m_\mu+m_\tau)/ {(\sqrt{m}_e+\sqrt{m}_\mu
+\sqrt{m}_\tau)^2}$ and
$r_{123}=\sqrt{m_e m_\mu m_\tau}/{(\sqrt{m}_e+\sqrt{m}_\mu
+\sqrt{m}_\tau)^3}$ are given by
Eqs.(2.21) and (2.23), respectively.
The deviation parameter $\eta$ from the bilinear form
$\langle\Phi_d\rangle^2$ in the down-quark mass formula
(2.8) is given by (2.22).
Those observable quantities are described by the parameters
$\alpha$, $\beta$, $\gamma$ and $\delta$ which are invariant 
under the scale transformations $\Phi \rightarrow \xi_\phi \Phi$,
$Y_e \rightarrow \xi_e Y_e$ and $Y_d \rightarrow \xi_d Y_d$.
Since those relations have been derived from the SUSY vacuum
conditions at a high energy scale, those do not suffer effects of 
a soft SUSY breaking based on a conventional scenario
at $\mu \sim 1$ TeV.

In order to reduce the number of the parameters, we have assumed
that the ratio $R$ is independent of the parameter 
$\gamma=\tilde{m}_{d\phi}/\tilde{m}_{dd}$ and we have gotten the constraint
$\gamma=\beta +1$.
Then, we have obtained a simple expression of $R$,
$R=1-4(1+\alpha)/9$, so that the ratio $R$ is given only by
the parameter $\alpha=\tilde{m}_{dd}/\tilde{m}_{ee}$.

In order to predict the charged lepton and down-quark masses,
we need explicit values of the parameters $\alpha$, $\beta$,
$\gamma$ and $\delta$.  
In Sec.3, we have demonstrated a trial scenario to obtain 
those parameters.
By assuming that these parameters have rational relations,
we have obtained $\alpha=-1/4$ and $\beta=-2$.
The result $\alpha=-1/4$ gives the desirable relation
$R=(m_e+m_\mu+m_\tau)/ {(\sqrt{m}_e+\sqrt{m}_\mu
+\sqrt{m}_\tau)^2}=2/3$, while the value $\beta=-2$ gives
$\eta\simeq -0.20$, which is somewhat deviated from the 
value $\eta \simeq -0.12$ estimated from the observed values
of $m_d/m_s$ and $m_s/m_b$.
In order to fit the ratio $r_{123}=\sqrt{m_e m_\mu m_\tau}/{
(\sqrt{m}_e+\sqrt{m}_\mu+\sqrt{m}_\tau)^3}$, a concise 
rational value $\delta=1/3$ is required. 
Such a value $\delta=1/3$  seems to be plausible.
However, the scenario given in Sec.3 is highly speculative, 
so that it should not be taken seriously.
How to get those parameter values more naturally is a future
task to us.
However, as we have demonstrated in Sec.3, the idea that
those parameter values are described by simple and rational
numbers seems to be promising.

By the way, in the present model, we have assumed that
the nonet field $\Phi$ is Hermitian.
Therefore, the parameters in the superpotential have been
taken real.
This does not mean that all components of $\langle\Phi\rangle$ 
are real, although three eigenvalues are real.
Therefore, the model can include a $CP$-violating phase.
However, in order to give such a phase explicitly, we will
need further modification to the superpotential form (2.1).

In the present paper, we have not investigated the up-quark 
and neutrino mass matrices.
We suppose that the up-quark and neutrino mass spectra are 
given by a similar mechanism for another nonet field $\Phi_u$.
(There is a possibility that the observed up-quark and 
neutrino (Dirac) mass spectra are also described by the forms
$m_{ui} \propto (z_{ui})^2 +\eta z_{ui}$ and
$m_{\nu i}^{Dirac} \propto (z_{ui})^2$ if we assume a VEV spectrum
$z_{ui} \delta_{ij} =\langle (\Phi_u)_{ij}\rangle
/\sqrt{{\rm Tr}[\Phi_u\Phi_u]}$ different from 
$\langle\Phi_d\rangle$.)
However, the present formulation is applicable only to 
mass spectra.
In order to give non-trivial flavor mixings, the diagonal
bases of $\langle\Phi_u\rangle$ and $\langle\Phi_d\rangle$
must be different from each other.
If we have a superpotential $W$ which consists of two sets
$W_u(\Phi_u, Y_u, Y_\nu)$ and $W_d(\Phi_d, Y_d, Y_e)$ and
in which there are no cross terms between $(\Phi_u, Y_u, Y_\nu)$
and $(\Phi_d, Y_d, Y_e)$, we can take two different bases,
$\langle \Phi_u\rangle$-diagonal basis and 
$\langle \Phi_d\rangle$-diagonal basis, separately. 
However, since, in such a model, there are no parameters
which describe relations between  $\langle \Phi_u\rangle$ and
$\langle \Phi_d\rangle$, we cannot predict the mixing matrices.
As we stressed in Sec.1, the most distinctive feature of the
present model is that the scenario does not include any
explicit flavor symmetry breaking parameter.
However, in order to give an explicit relation between  
$\langle \Phi_u\rangle$ and $\langle \Phi_d\rangle$,
we will be obliged to introduce some symmetry breaking term
as a flavor-basis fixing term (for example, see 
Ref.\cite{Haba-Koide07}).
 
The idea that the flavor mass spectra originate in a
VEV structure of a U(3)-nonet scalar seems to be promising 
for unified understanding of quark and lepton masses and mixings.

\vspace{6mm}

\centerline{\large\bf Acknowledgments} 
This work is an offshoot of a series of works in
collaboration with N.~Haba.
The author would like to thank N.~Haba for helpful
conversations.
This work is supported by the Grant-in-Aid for
Scientific Research, Ministry of Education, Science and 
Culture, Japan (No.18540284).



\end{document}